\documentclass[aps, pra, reprint, lengthcheck, footinbib, showpacs,  citeautoscript]{revtex4-2}

\usepackage[T1]{fontenc}

\usepackage{multirow}

\usepackage{amsmath}

\usepackage{amsmath}

\usepackage{braket}
\usepackage{graphicx}

\usepackage[colorlinks, urlcolor=blue, citecolor=blue, linkcolor=blue, pdfstartview=FitH]{hyperref}
\usepackage[all]{hypcap}

\usepackage{dcolumn}
\usepackage{bm}
\usepackage{epstopdf}

\usepackage{xcolor}
\usepackage[normalem]{ulem}
\definecolor{color}{rgb}{0.11,0.45,0.02}

\begin{document}

\title{Land\'e $g$-factors of electrons and holes strongly confined in CsPbI$_3$ perovskite nanocrystals in glass}

\author{Sergey~R.~Meliakov$^{1}$, Evgeny~A.~Zhukov$^{2,1}$, Vasilii~V.~Belykh$^{2}$, Mikhail~O.~Nestoklon$^{2}$,  Elena~V.~Kolobkova$^{3,4}$, Maria~S.~Kuznetsova$^5$, Manfred~Bayer$^{2}$, Dmitri~R.~Yakovlev$^{2,1}$}

\affiliation{$^{1}$P.N. Lebedev Physical Institute of the Russian Academy of Sciences, 119991 Moscow, Russia}
\affiliation{$^{2}$Experimentelle Physik 2, Technische Universit\"at Dortmund, 44227 Dortmund, Germany}
\affiliation{$^{3}$ITMO University, 199034 St. Petersburg, Russia}
\affiliation{$^{4}$St. Petersburg State Institute of Technology, 190013 St. Petersburg, Russia}
\affiliation{$^{5}$Spin Optics Laboratory, St. Petersburg State University, 198504 St. Petersburg, Russia}

\date{\today}

\begin{abstract}
The Land\'e $g$-factor of charge carriers is a key parameter in spin physics controlling spin polarization and spin dynamics. In turn, it delivers information of the electronic band structure in vicinity of the band gap and its modification in nanocrystals provided by strong carrier confinement. The coherent spin dynamics of electrons and holes are investigated in CsPbI$_3$ perovskite nanocrystals with sizes varied from 4 to 16 nm by means of time-resolved Faraday ellipticity at the temperature of 6~K. The Land\'e $g$-factors of the charge carriers are evaluated through the Larmor spin precession in magnetic fields up to 430~mT across the spectral range from 1.69 to 2.25~eV, provided by variation of the nanocrystal size. The spectral dependence of the electron $g$-factor follows the model predictions when accounting for the mixing of the electronic bands with increasing confinement resulting from a decrease of the nanocrystal size. The spectral dependence of the hole $g$-factor, changing from $-0.19$ to $+1.69$, is considerably stronger than expected from the model. We analyze several mechanisms and conclude that none of them can be responsible for this difference. The renormalizations of the electron and hole $g$-factors roughly compensate each other, providing spectral independence for the bright exciton $g$-factor with a value of about $+2.2$.
\end{abstract}

\maketitle


\section{Introduction}

Colloidal nanocrystals (NCs) made of lead halide perovskite semiconductors provide attractiveness for applications in photovoltaics, optoelectronics, electronics, and beyond~\cite{Kovalenko2017,Shamsi2019,Dey2021,Mu2023,Huang2023}. They are commonly synthesized by colloidal chemistry in solution and can be composed of hybrid organic-inorganic or fully-inorganic materials. The fully-inorganic NCs made, e.g., of CsPbI$_3$, CsPbBr$_3$, or CsPbCl$_3$, show considerably higher stability in ambient conditions compared with the materials containing organic components. Encapsulation of the NCs allows one to further enhance their stability. One of the successful approaches in this respect is to synthesize the perovskite NCs in a glass matrix from a melt~\cite{Li2017,Liu2018,Liu2018a,Ye2019,Kolobkova2021,Belykh2022}. The surface of such NCs is free from organic ligands that are typical for the growth in solution. Also, the glass samples can be easily polished to get surfaces of optical quality, which facilitates their optoelectronic application.

The electronic band structure of lead halide perovskite NCs considerably differs from common III-V and II-VI semiconductor NCs and quantum dots~\cite{Efros2021,kirstein2022nc,nestoklon2023_nl}. Most importantly, the band gap is formed by conduction band and valence band states with a simple structure with spin $1/2$, which results in different spin level structure of the bright and dark exciton states, as compared with III-V and II-VI NCs~\cite{Belykh2022}. An in-depth understanding of the physical properties of perovskite NCs calls for a comprehensive study by various techniques, among which optical techniques are particularly favorable and informative. 

Spin-dependent phenomena typically are helpful, as they deliver information on the charge carriers interacting with each other, with the crystal lattice, and with the nuclear spin system~\cite{Vardeny2022_book}. The spin polarization and spin dynamics are sensitive to the details of the band structure and crystal symmetry. The spin properties of perovskite NCs have been addressed by several optical and magneto-optical techniques, namely, optical orientation and optical alignment~\cite{Nestoklon2018}, polarized emission in magnetic field~\cite{canneson2017}, time-resolved Faraday/Kerr rotation~\cite{Crane2020,Grigoryev2021,Lin2022,Meliakov2023NCs,kirstein2023_SML}, time-resolved differential transmission~\cite{Strohmair2020,Han2022,Gao2024}, and optically-detected nuclear magnetic resonance~\cite{kirstein2023_SML}. Among them, time-resolved Faraday/Kerr rotation is particularly informative, as it provides data on the electron and hole Land\'e $g$-factors, spin coherence, spin dephasing, and longitudinal spin relaxation.  Carrier spin coherence up to room temperature has been reported in CsPbBr$_3$ NCs~\cite{Crane2020,Meliakov2023NCs} and its optical manipulation has been demonstrated~\cite{Lin2022,Zhu2024}. 

The Land\'e $g$-factor of charge carriers and excitons determines their Zeeman splitting in magnetic field and thus is a key parameter in spin physics. We found recently that in  bulk lead halide perovskites the electron and hole as well as exciton $g$-factors follow universal dependences on the band gap energy~\cite{kirstein2022nc,Kopteva_gX_2023}. In NCs, the carrier confinement induces mixing of the electronic ground states with higher bands, which results in a considerable renormalization of the electron $g$-factor ($g_e$), as predicted theoretically and found experimentally in  CsPbI$_3$ NCs in glass~\cite{nestoklon2023_nl,Meliakov2024paper1}. For the hole $g$-factor ($g_h$) the predicted renormalization is small, which is in line with the experimental data for relatively large NCs with sizes of $8-16$~nm~\cite{nestoklon2023_nl}. To complete the picture, it is important to study the $g$-factors in smaller NCs with strong geometric confinement of the carriers. Especially, recently an unusually strong spectral dependence of $g_h$ in CsPbBr$_3$ NCs, which considerably exceeds the model expectations, was found~\cite{Meliakov2024paper3}.    

In this paper, we study the coherent spin dynamics of electrons and holes in CsPbI$_3$ NCs with sizes in the range $4-16$~nm, embedded in a glass matrix, by time-resolved Faraday ellipticity. The spectral dependences of the electron and hole $g$-factors are measured in the large range of exciton energies of $1.69-2.25$~eV accessible by varying carrier confinement. Both $g$-factors become considerably renormalized with increasing carrier confinement (i.e., decreasing NC size). The changes of the electron $g$-factor are in line with model predictions, while the hole $g$-factor demonstrates a much larger renormalization, which mechanism needs to be understood.

\section{Experimental results}

We study experimentally a set of CsPbI$_3$ NCs embedded in a fluorophosphate glass matrix, namely, six samples with different NC sizes covering the range of  $4-16$~nm. They are labeled as samples \#1, \#2, \#3, \#4, \#5, and \#6 with decreasing NC size with increasing sample number. The samples show inhomogeneous broadening of their optical properties, provided by the considerable dispersion of NC size within one sample. Therefore, in our study we rely on the energy of exciton resonance as a characteristic of the carrier confinement energy. In the studied set of NCs the exciton energy covers the range of $1.69-2.25$~eV.  Detailed information on the optical properties and carrier $g$-factors for the larger NCs (samples \#1, \#2 and \#3 with sizes in range of $8-16$~nm) can be found in Refs.~\onlinecite{nestoklon2023_nl,Meliakov2024paper1}.   

\begin{figure*}[hbt!]
\includegraphics[width=1.7\columnwidth]{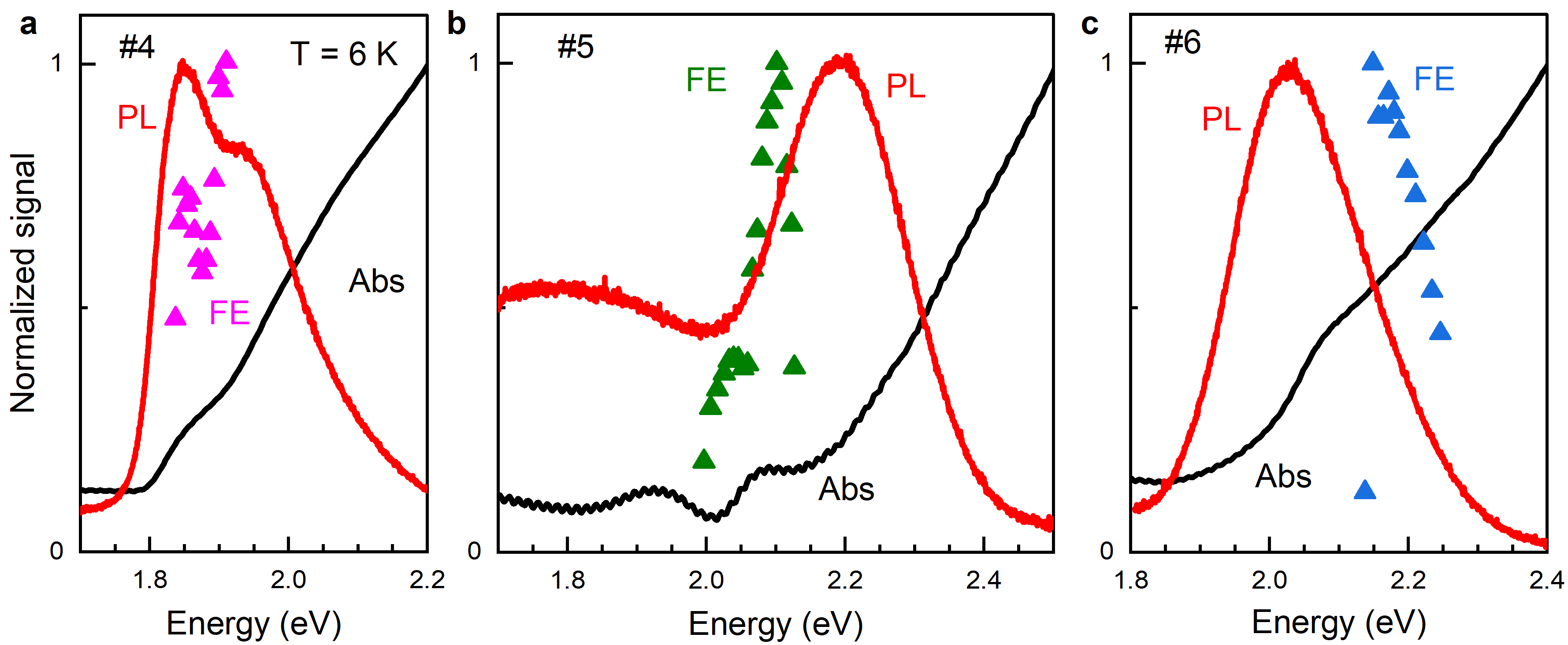}
\caption{Spectral dependences of the photoluminescence intensity (red line), absorption (black line), and Faraday ellipticity amplitude (symbols) in CsPbI$_3$ NCs (samples  \#4, \#5, and \#6), measured at the temperature of 6~K. 
}
\label{fig:Samples}
\end{figure*}

Figure~\ref{fig:Samples} shows absorption and photoluminescence (PL) spectra measured at the temperature of $T=6$~K for the samples \#4, \#5, and \#6 in the size range of $4-8$~nm. The PL lines have a typical half width at half maximum of about 200~meV. In the time-resolved Faraday ellipticity (TRFE) experiments, the laser is tuned spectrally, which allows us to address NCs with a specific exciton transition energy corresponding to a specific NC size. For that, spectrally narrow laser pulses with 1~meV width and 1.5~ps duration are used. The corresponding amplitude of the TRFE signals is shown in Figure~\ref{fig:Samples} by the symbols. 

TRFE is used to measure the coherent dynamics of the electron and hole spins in a magnetic field. This pump-probe-based technique exploits polarized laser pulses~\cite{Yakovlev_Ch6,belykh2019}: spin-oriented carriers are photogenerated by the circularly-polarized pump pulses and the dynamics of their spin polarization are detected through the change of the ellipticity of the linearly polarized probe pulses~\cite{YugovaPRB09,Glazov2010}. We perform these experiments at the temperature of $T=6$~K in magnetic fields, $B$, up to 430~mT, applied in the Voigt geometry, i.e., perpendicular to the light wave vector.

\begin{figure*}[hbt!]
\centering
\includegraphics[width=1.6\columnwidth]{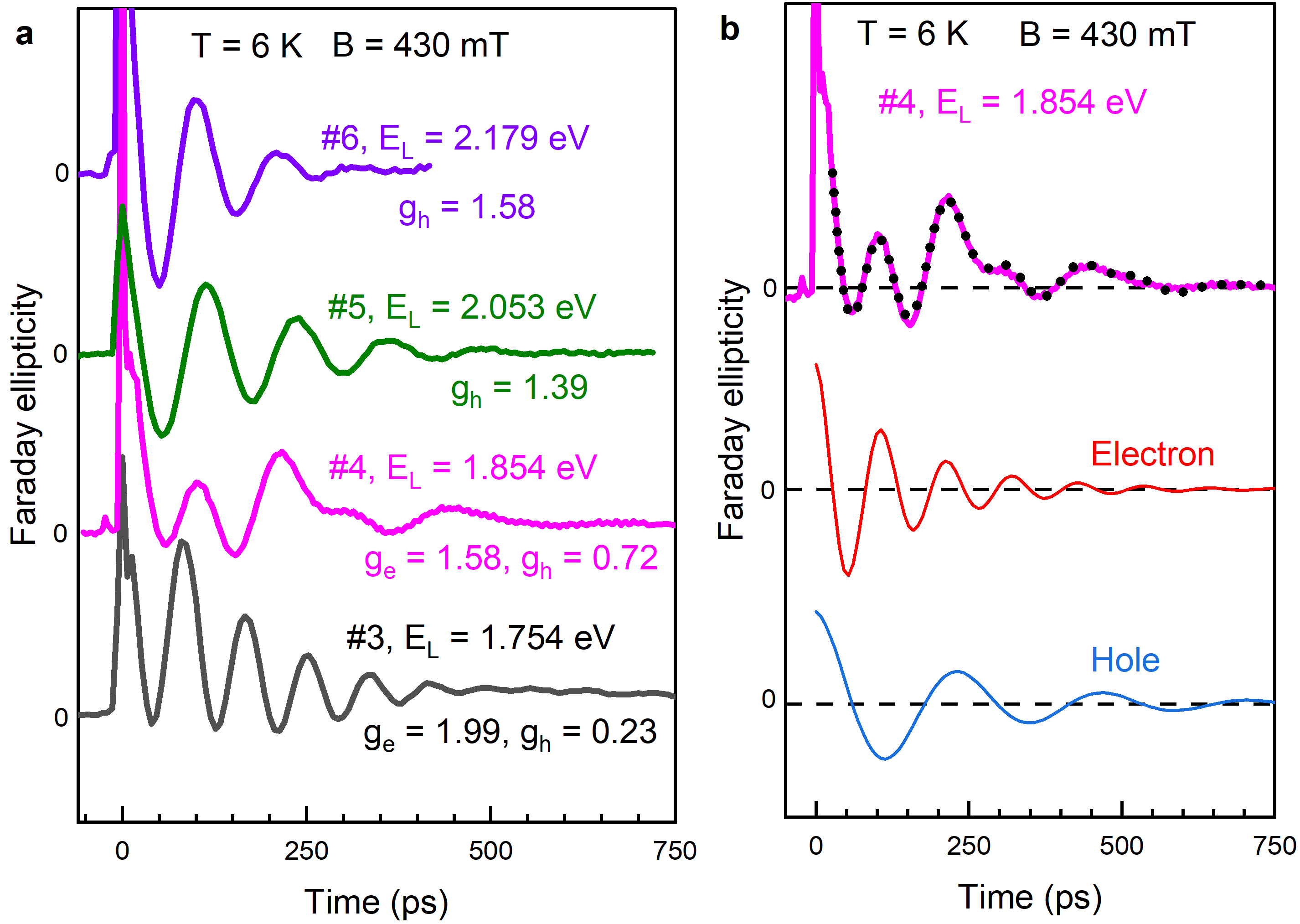}
\caption{(a) Spin dynamics in the CsPbI$_3$ NCs (samples \#3, \#4, \#5 and \#6) in 430~mT magnetic field. 
(b) The upper graph shows the FE dynamics in sample \#4. Black dotted line is a fit with eq.~\eqref{eq:Voigt}. The two lower graphs show the oscillatory components in the spin dynamics, corresponding to the electron (red line, $|g_e|=1.58$ and $T_\text{2,e}^*=150$~ps) and the hole (blue line, $|g_h|=0.72$ and $T_\text{2,h}^*=220$~ps).
}
\label{fig:Dynamics}
\end{figure*}

The TRFE dynamics measured in the samples \#3, \#4, \#5, and \#6 at $B=430$~mT are shown in Figure~\ref{fig:Dynamics}a. The spin dynamics demonstrate oscillating signals due to the Larmor spin precession of carrier spins about external magnetic field~\cite{Grigoryev2021,Meliakov2024paper1}. The Larmor precession frequency, $\omega_\text{L}$, is determined by the $g$-factor and scales with the magnetic field $B$ according to:
\begin{equation}
\omega_\text{L} = |g|{{\mu}_\text{B}}B/{\hbar} \,.
\label{eq:LarmFreq}
\end{equation}
Here ${\mu}_\text{B}$ is the Bohr magneton and $\hbar$ is the reduced Planck constant. The decay of the oscillations is described by the spin dephasing time $T_2^*$, which at low temperatures mainly reflects the inhomogeneous spread of the Larmor precession frequencies provided by the spread of $g$-factors, $\Delta g$. At weak magnetic fields down to zero field strength, the hyperfine interaction of carriers with the nuclear spin fluctuations also contributes to the carrier spin dephasing~\cite{Meliakov2024paper3}.

One can see in Figure~\ref{fig:Dynamics}a that the spin dynamics in samples \#3 and \#4 have two oscillating components. To identify them and to extract their parameters we fit the dynamics with the following function:
\begin{equation}
A_{\rm FE}(t) = \sum\limits_{i=\text{e,h}} S_{\text{0},i} \cos(\omega_{\text{L},i}t) \exp(-t/{T_{\text{2},i}^*}) \,.
\label{eq:Voigt}
\end{equation}
Here, $S_\text{0,e}$ and $S_\text{0,h}$ are the initial light-induced spin polarizations of electrons and holes, respectively. $T_\text{2,e}^*$ and $T_\text{2,h}^*$ are the electron and hole spin dephasing times. An example of such a fit is shown by the dotted line in Figure~\ref{fig:Dynamics}b for the sample \#4. There, also the separate components of the electron and hole to the spin dynamics are given. According to Ref.~\onlinecite{nestoklon2023_nl}, the oscillations with larger ($\omega_\text{L,e}=58.5$~rad/ns) and smaller ($\omega_\text{L,h}=26.5$~rad/ns) Larmor frequencies correspond to the electron and hole spin precession, respectively. The extracted spin dephasing times at $B=430$~mT are $T_\text{2,e}^*=150$~ps and $T_\text{2,h}^*=220$~ps. A similar phenomenology is observed for the sample \#3. The electron and hole contributions could be also isolated in the spin dynamics of samples \#1, \#2 and \#3, see Refs.~\onlinecite{nestoklon2023_nl,Meliakov2024paper1}. For the samples \#5 and \#6, only one oscillating component can be detected, which we assign to the holes based on its spectral dependence, as discussed below.

\begin{figure*}[hbt!]
\centering
\includegraphics[width=2\columnwidth]{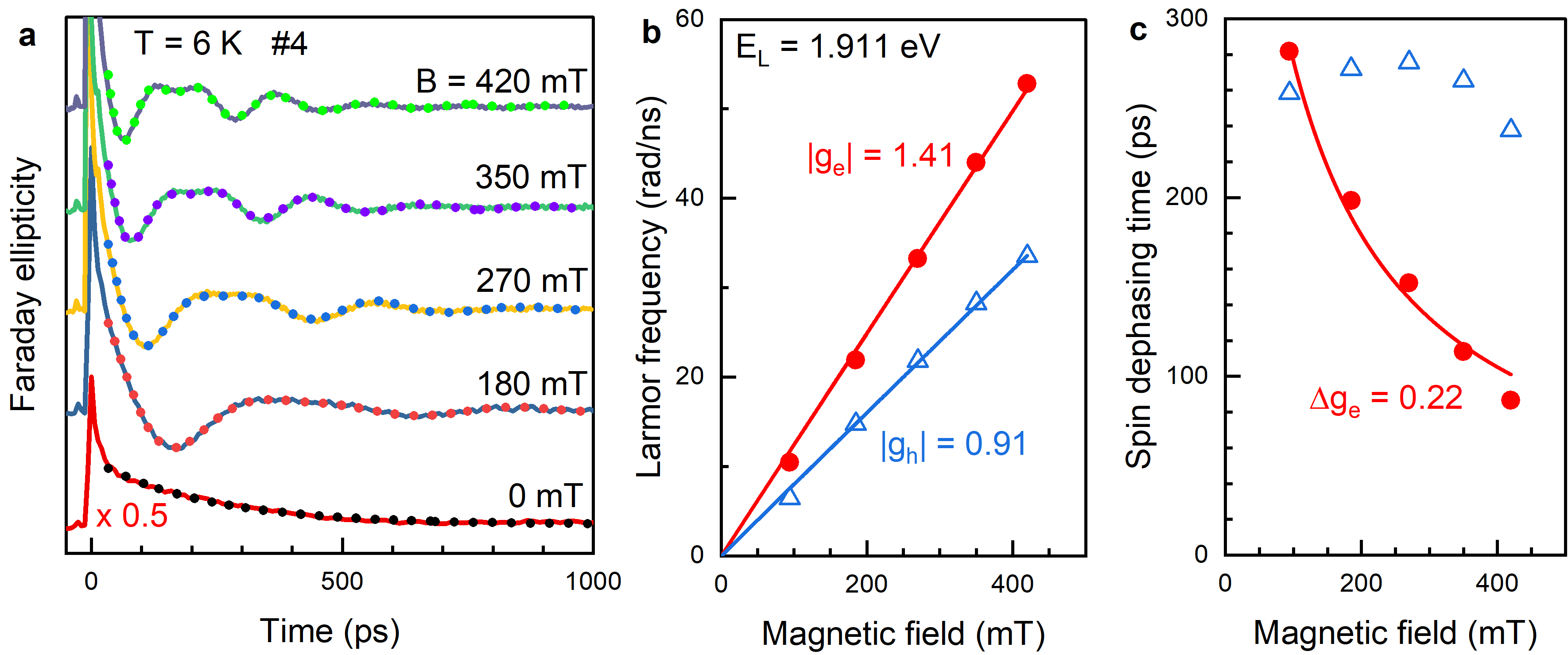}
\caption{Magnetic field dependence of the spin dynamics in CsPbI$_3$ NCs of sample \#4, measured at $E_\text{L}=1.911$~eV.
(a) TRFE dynamics in various Voigt magnetic fields. Dotted lines show fits to the experimental data using eq.~\eqref{eq:Voigt}.
(b) Magnetic field dependences of the electron (red circles) and hole (blue triangles) Larmor precession frequencies. Lines show linear fits using eq.~\eqref{eq:LarmFreq}.
(c) Magnetic field dependences of the electron (red circles) and hole (blue triangles) spin dephasing time $T_2^*$. Red line shows fit using eq.~\eqref{eq:InhDeph}.
}
\label{fig:Magnetic} 
\end{figure*}

We measure the spin dynamics for all samples in various magnetic fields. An example of the sample \#4, measured at the energy of $1.911$~eV, is shown in Figure~\ref{fig:Magnetic}a. At zero magnetic field, the decay is close to single exponential with the decay time of 230~ps. With increasing magnetic field, the Larmor spin precession signal becomes pronounced. The magnetic field dependences of the Larmor precession frequencies are shown in Figure~\ref{fig:Magnetic}b. As expected, they scale linearly with the magnetic field strength and show no offset when extrapolated to zero field. This confirms that the spin signals originate from  resident carriers, i.e., from NCs charged either by an electron or by a hole~\cite{Grigoryev2021}, rather than from excitons. In the latter case electron-hole exchange interaction would lead to exciton spin beats even at zero magnetic field~\cite{Han2022,Gao2024,Zhu2024}. The resident carriers in the NCs can appear from long-living photocharging, where either the electron or the hole from a photogenerated electron-hole pair escapes from the NC. As a result, a fraction of NCs in the ensemble is charged with electrons, another fraction of NCs is charged with holes, while the rest NCs remain neutral. According to eq.~\eqref{eq:LarmFreq}, the slopes of the dependences in Fig.~\ref{fig:Magnetic}(b) correspond to the absolute values of the electron and hole $g$-factors: $|g_{\rm e}| = 1.41$ and $|g_{\rm h}| = 0.91$. It was shown in Refs.~\onlinecite{nestoklon2023_nl,Meliakov2024paper1} that in CsPbI$_3$ NCs the electron and hole $g$-factors are positive in this energy range. 

The magnetic field dependences of the spin dephasing times of electrons ($T_\text{2,e}^*$) and holes ($T_\text{2,h}^*$) are shown in Figure~\ref{fig:Magnetic}c. The spin dephasing time $T_\text{2,h}^*$ barely depends on magnetic field amounting to about $260$~ps, while $T_\text{2,e}^*$  decreases from 280~ps to 90~ps with the magnetic field growing from 90~mT to 420~mT. This behavior is typical for inhomogeneous spin ensembles with a finite $g$-factor spread $\Delta g$ and can be described by the following expression~\cite{Yakovlev_Ch6}:
\begin{equation}
\frac{1}{T_2^*(B)} {\approx} \frac{1}{{T_2^*}(0)} + \frac{{{\Delta}g}{{\mu}_\text{B}}B}{\hbar}.
\label{eq:InhDeph}
\end{equation}
Here, ${T_2^*}(0)$ is the spin dephasing time at zero magnetic field. Fitting the experimental data with eq.~\eqref{eq:InhDeph}, yields $\Delta g_{\text{e}}=0.22$ for the electrons (the relative spread is $\Delta g_{\text{e}} / g_{\text{e}}=16\%$). In Ref.~\onlinecite{Meliakov2024paper1} we reported $\Delta g_{\text{e}}=0.25$ for the electrons and $\Delta g_{\text{h}}=0.07$ for the holes in sample \#1.

\begin{figure}[hbt!]
\includegraphics[width=1\columnwidth]{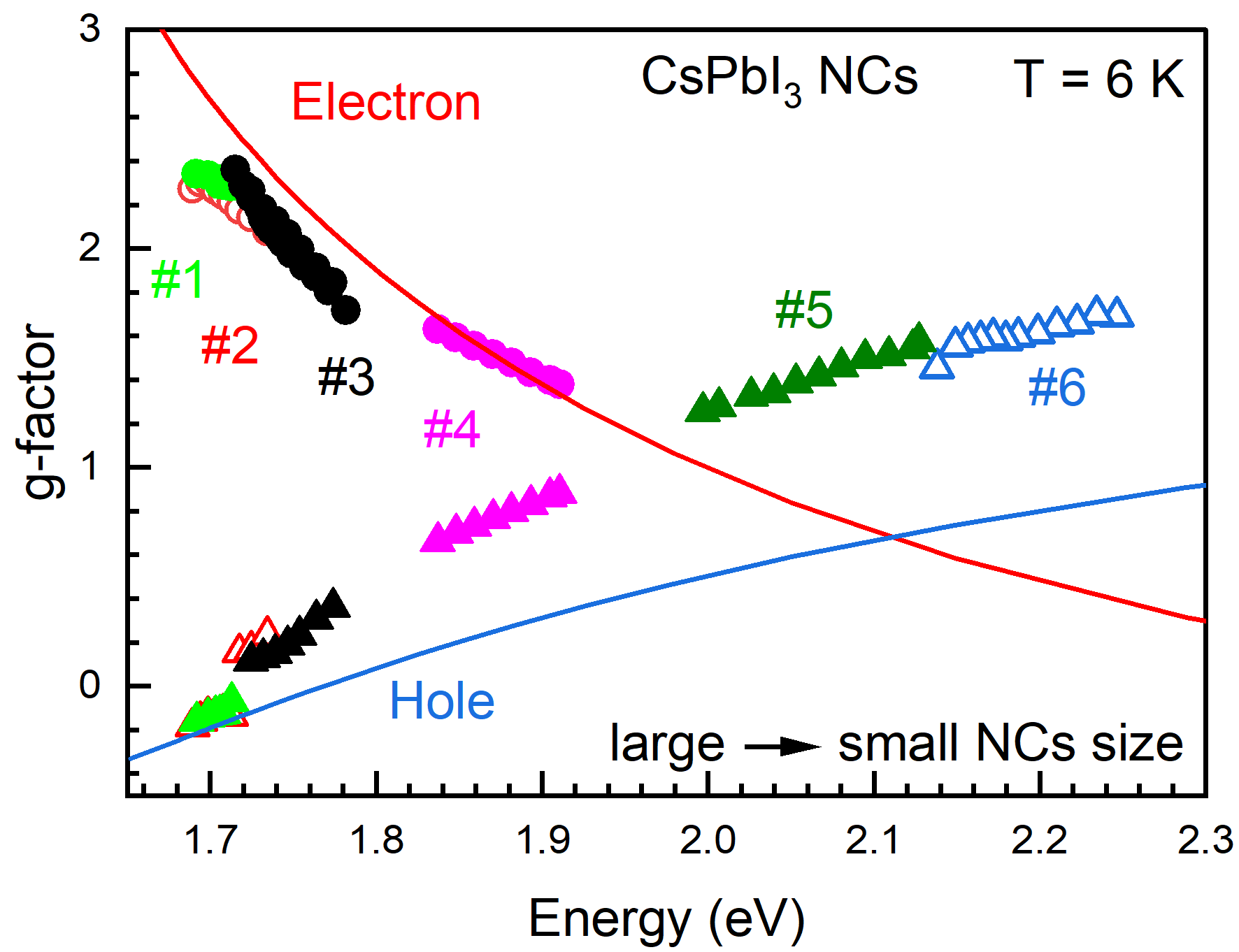}
\caption{Spectral dependences of the electron (circles) and hole (triagles) $g$-factors in CsPbI$_3$ NCs in glass: samples \#1 (light green symbols), \#2 (red), \#3 (black), \#4 (pink), \#5 (dark green), and \#6 (blue). Solid lines give the calculations for CsPbI$_3$ NCs from Ref.~\onlinecite{nestoklon2023_nl}.
}
\label{fig:g}
\end{figure}

The spectral dependences of the electron and hole $g$-factors for the studied CsPbI$_3$ NCs are shown in Figure~\ref{fig:g}. The electron $g$-factors are measured for samples \#1 to \#4. They decrease from $+2.3$ to $+1.4$ with energy increasing from 1.69~eV to 1.91~eV. This dependence closely follows the theory predictions shown by the red line, where the used model accounts for mixing of the electron ground states with higher bands when the confinement energy increases. The situation is different for the hole $g$-factor. It changes from $-0.19$ to $+1.69$ for the energy increasing from 1.69~eV to 2.25~eV. The renormalization of the $g$-factor  is much stronger than the model predictions shown by the blue line in Figure~\ref{fig:g}. The difference between experimental and theoretical $g$-factors reaches 0.8 for the smallest NCs. In fact, the model considerations reported in Ref.~\onlinecite{nestoklon2023_nl} conclude on a rather weak effect of the hole quantum confinement on its $g$-factor, and the blue line in Figure~\ref{fig:g} is close to the band gap dependence of $g_h$ in bulk crystals, see Ref.~\onlinecite{kirstein2022nc}. 

\begin{figure}[hbt!]
\includegraphics[width=1\columnwidth]{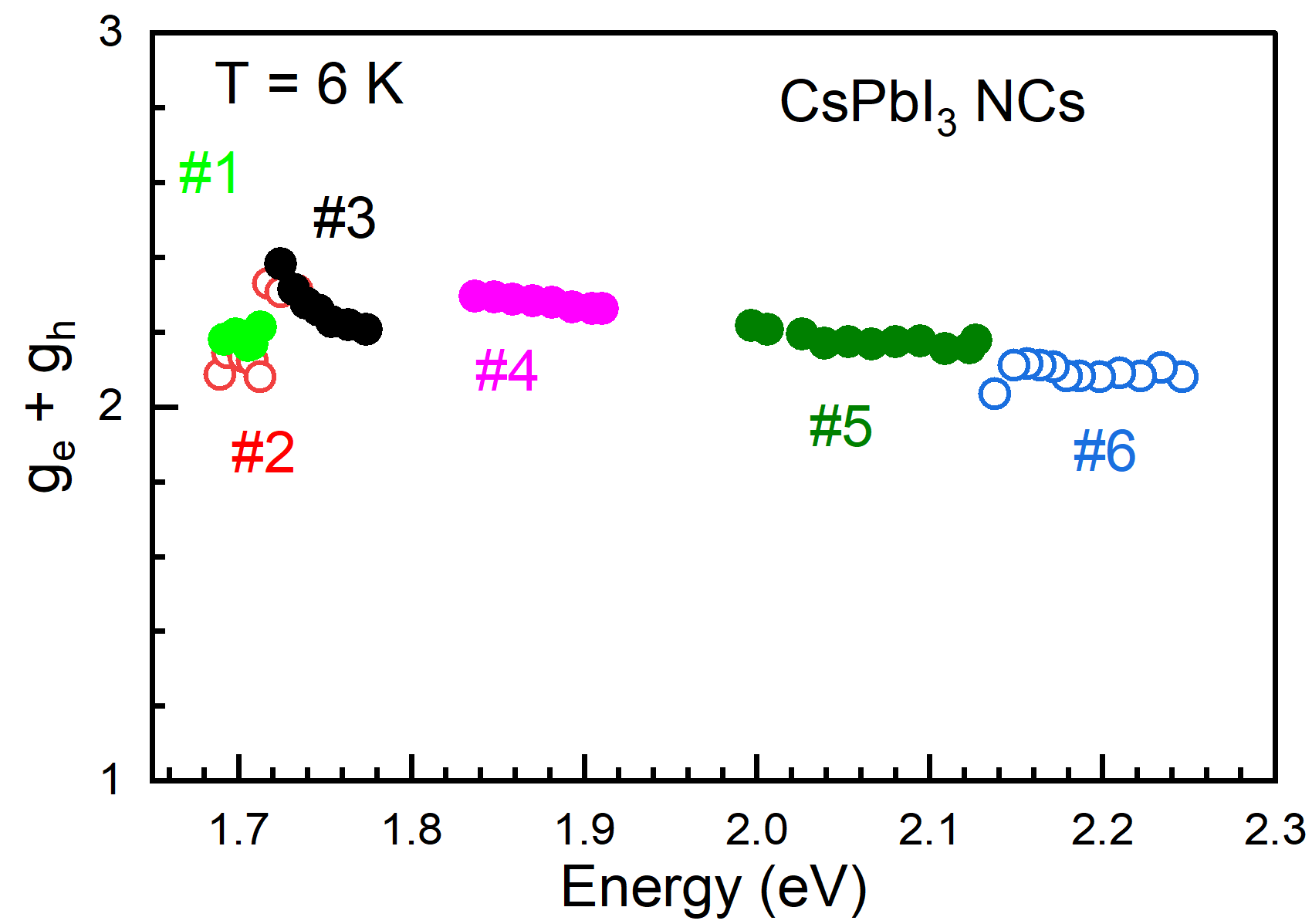}
\caption{Spectral dependence of the bright exciton $g$-factor in CsPbI$_3$ NCs, calculated from the electron and hole $g$-factors as  $g_{\rm BX}=g_{\rm e}+g_{\rm h}$. For the samples \#1 to \#4, the experimental values for  $g_{\rm e}$ and $g_{\rm h}$ are taken. For the samples \#5 and \#6, we take the experimental values of $g_{\rm h}$ and the modeled values of $g_{\rm e}$ from Ref.~\onlinecite{nestoklon2023_nl}. 
}
\label{fig:gX}
\end{figure}

It is instructive to analyze the spectral dependence of the bright exciton $g$-factor, which is composed of the electron and the hole $g$-factor. In lead halide perovskite NCs the bright exciton $g$-factor is given by $g_{\rm BX}=g_{\rm e}+g_{\rm h}$. We plot it in Figure~\ref{fig:gX} by taking the experimental data from Figure~\ref{fig:g}. For the samples \#5 and \#6, for which experimental results for $g_{\rm e}$ are not available, we take the modeled values from Ref.~\onlinecite{nestoklon2023_nl}, which are shown by the red line in Figure~\ref{fig:g}.  One can see in Figure~\ref{fig:gX} that surprisingly $g_{\rm BX}$ is almost independent of energy in the whole range of $1.69-2.25$~eV, amounting to $g_{\rm BX} \approx +2.2$. Note that in bulk crystals of the lead halide perovskites with different compositions the bright exciton $g$-factor is also rather constant as it remains in the range of $+2.3$ to $+2.7$ for variation of the band gap from 1.5 to 3.2~eV~\cite{Kopteva_gX_2023}. Recently, the exciton $g$-factor value of $|g_{\rm BX}| = 2.29$ was reported for CsPbI$_3$ NCs grown in solution~\cite{Gao2024}. It was measured at $T=200$~K for NC size of about 7.2~nm (at energy of 1.812~eV) from the exciton spin beats in time-resolved differential absorption. This value is in a very good agreement with our results for CsPbI$_3$ NCs in glass shown in Figure~\ref{fig:gX}.

\section{Discussion}

Let us analyze the factors that contribute to the spectral dependence of the electron and hole $g$-factors in CsPbI$_3$ NCs. As a starting point we take the universal dependence of the carrier $g$-factors on the band gap found experimentally for cryogenic temperatures and confirmed theoretically for lead halide perovskite bulk crystals \cite{kirstein2022nc}. These studies were recently extended to address the role of quantum confinement of charge carriers in NCs on the $g$-factors \cite{nestoklon2023_nl}. 

In the atomistic calculations using the empirical tight-binding method (ETB) based on density functional theory (DFT) calculations~\cite{Nestoklon2021}, the hole $g$-factor as function of energy is expected to follow the bulk trend, while for electrons the situation is more complicated~\cite{nestoklon2023_nl}. The electron $g$-factor demonstrates a strong effect of the quantum confinement, decreasing significantly with an increase of the effective band gap (a decrease of the NC size) and even changing sign from positive to negative. The calculations may be explained qualitatively in the framework of the {\bf k}$\cdot${\bf p} model~\cite{kirstein2022nc} extended to account for the geometric confinement~\cite{nestoklon2023_nl}. The strong renormalization of the electron $g$-factor originates from the mixing of the bottom conduction band with the spin-orbit split electron states due to this confinement.  In Ref.~\onlinecite{nestoklon2023_nl} the experimental data were limited, but the expected trend for the electron $g$-factor was quite pronounced. In the present paper, the experimental results cover a broader energy range. As seen from Figure~\ref{fig:g}, the experimentally measured electron $g$-factors perfectly agree with the theoretical predictions.

In contrast, the hole $g$-factors measured across the wide energy range deviate significantly from the atomistic calculations and the {\bf k}$\cdot${\bf p} analysis. While the qualitative trend is in agreement with the predictions, for large quantum confinement energies the experimental data almost twice exceed the predicted values, see Figure~\ref{fig:g}. Let us discuss the possible sources of this discrepancy. 

One of the complications in comparing theory with experiment is the uncertainty in NC quantum confinement. On the theory side, the transition energy is the difference between the energies of the first quantum confined levels of the electron in the conduction band and the hole in the valence band. On the experimental side, the transition energy is the photon energy of the laser exciting the TRFE signal. The difference between these energies for the same size NCs (due to the exciton binding energy, dielectric screening, etc.) is expected to be small but not zero and, more importantly, may change with the NC size. However, the difference for the hole $g$-factors in Figure~\ref{fig:g} cannot be explained by this effect: the $g$-factor is not sensitive enough to energy, and to explain the observed difference an unrealistic renormalization of the energy scale would have to be assumed. 

A second possible source of the deviation is the non-parabolicity of the hole energy dispersion, which could result in an overestimated hole quantum confinement energy in theory. We believe that this also cannot explain the observed difference as the non-parabolicity of the dispersion (both for electron and hole) is carefully taken into account in the ETB calculations. The electron and hole energy dispersions in the ETB accurately reproduce the DFT results, see the Supporting Information of Ref.~\onlinecite{nestoklon2023_nl}, and the results qualitatively agree with the {\bf k}$\cdot${\bf p} calculations. This means that even if non-parabolicity is seriously underestimated in theory, its correct consideration will hardly compensate for the difference between experiment and theory. 

The most probable cause of the discrepancy is the mixing of the upper valence band with halide bands lying approximately $1.5$~eV below the top of valence band. This mixing is underestimated in the ETB consideration. The nearest-neighbor empirical tight-binding calculations allow one to describe the upper valence and three lowest electron bands with meV precision \cite{Nestoklon2021}. However, the correct dispersion in the lower valence bands formed from the $p$-orbitals of halide atoms demands for accounting the second-neighbor interaction \cite{Kashikar2018}, which is missing in the scheme used in Ref.~\onlinecite{nestoklon2023_nl}. If the admixture of the lower valence band and/or the effect of the spin-orbit interaction on its dispersion is underestimated, the renormalization of the hole $g$-factor due to quantum confinement may also be significantly underestimated. An additional analysis of the theoretical description, which goes beyond the present study, is needed to reveal whether this factor can close the gap between theory and experiment. 

\section{Conclusions}

We have measured the coherent spin dynamics of electrons and holes in CsPbI$_3$ perovskite nanocrystals with sizes varying from 4 to 16 nm  by time-resolved Faraday ellipticity. We evaluate the Land\'e $g$-factors of charge carriers in the regime of strong quantum confinement providing a spectral shift of exciton resonance from 1.69 to 2.25~eV. The electron $g$-factor demonstrates a strong renormalization caused by the admixture of the upper conduction band to the ground state, in agreement with theoretical predictions. The spectral dependence of hole $g$-factor qualitatively follows the trend expected for bulk crystals, but also demonstrates a significant renormalization by a value similar to the changes of the electron $g$-factor, which cannot be explained by the theory. Interestingly, the renormalization of the electron and hole $g$-factors almost compensate each other, almost canceling the spectral dependence of the bright exciton $g$-factor leading to an almost constant value of $+2.2$. Understanding the details of the responsible mechanisms for this behavior would allow one to accurately refine the band parameters for the lead halide perovskites and their NCs.

\section{Experimental Section}


\textit{Samples}:  The studied CsPbI$_3$ nanocrystals embedded in fluorophosphate Ba(PO$_3$)$_2$-AlF$_3$ glass were synthesized by rapid cooling of a glass melt enriched with the components needed for the perovskite crystallization. Details of the method are given in Refs.~\onlinecite{Kolobkova2021,kirstein2023_SML}. The samples of fluorophosphate (FP) glass with the composition 35P$_2$O$_5$--35BaO--5AlF$_3$--10Ga$_2$O$_3$--10PbF$_2$--5Cs$_2$O (mol. \%) doped with BaI$_2$ were synthesized using the melt-quench technique. The glass synthesis was performed in a closed glassy carbon crucible at the temperature of $T=1050^\circ$C.

Six samples are investigated in this paper, which we label \#1, \#2, \#3, \#4, \#5, and \#6. Their technology codes are EK31, EK7, EK8, EK201, EK205 and EK203, respectively. Note that the first three samples (same codes and synthesis) were investigated in Refs.~\onlinecite{nestoklon2023_nl} and~\onlinecite{Meliakov2024paper1}.
About 50~g (for samples \#1, \#2, \#3) and 25~g (for samples \#4, \#5, \#6) of the batch were melted in the crucible for 20 -- 30~minutes, then the glass melt was cast on a glassy carbon plate, and pressed to form a plate with a thickness of about 2~mm. Samples with a diameter of 5~cm were annealed at the temperature of $50^\circ$C below $T_g=400^\circ$C to remove residual stresses. The CsPbI$_3$ perovskite NCs were formed from the glass melt during the quenching. The glasses obtained in this way are doped with CsPbI$_3$ NCs. The NC sizes in the initial glass were regulated by the concentration of iodide and the rate of cooling of the melt without heat treatment above $T_g$. They differ in the NC sizes, which is reflected by the relative spectral shifts of their optical spectra. The change of the NC size was achieved by changing the concentration of iodine in the melt. Due to the high volatility of iodine compounds and the low viscosity of the glass-forming fluorophosphate melt at elevated temperatures, an increase in the synthesis time leads to a gradual decrease of the iodine concentration in the equilibrium melt. Thus, it is possible to completely preserve the original composition and change only the concentration of iodine by means of a smooth change of the synthesis duration.


\textit{Time-resolved Faraday ellipticity}: To study the coherent spin dynamics of carriers we use a time-resolved pump-probe technique with detection of the Faraday ellipticity (TRFE)~\cite{Yakovlev_Ch6,Glazov2010}. Spin oriented electrons and holes are generated by circularly polarized pump pulses. The used laser system (Light Conversion) generates pulses of 1.5 ps duration with a spectral width of about 1~meV at the repetition rate of 25~kHz (repetition period 40~${\mu}$s). The laser photon energy is tuned in the spectral range of $1.65-2.30$~eV in order to resonantly excite NCs of various sizes. The laser beam is split into the pump and probe beams having the same photon energies. The time delay between the pump and probe pulses is controlled by a mechanical delay line. The pump beam is modulated with an electro-optical modulator between ${\sigma}^+$ and ${\sigma}^-$ circular polarization at the frequency of 26~kHz. The probe beam is linearly polarized. The Faraday ellipticity of the probe beam, which is proportional to the carrier spin polarization, is measured as function of the delay between the pump and probe pulses using a balanced photodetector, connected to a lock-in amplifier that is synchronized with the modulator. Both pump and probe beams have the power of 0.5~mW with spot sizes of about 100$~\mu$m. For the time-resolved  measurements the samples are placed in a helium-flow optical cryostat at $6$~K temperature. Magnetic fields up to 430~mT are applied perpendicular to the laser beam (Voigt geometry, $\textbf{B} \perp \textbf{k}$) using an electromagnet.

\section*{Conflicts of interest}
The authors declare no conflict of interest.

\section*{Data availability}
The data that support the findings of this study are available from the corresponding author upon reasonable request.

\section*{Acknowledgments}
The authors are thankful to M. M. Glazov for fruitful discussions. Research performed at the P. N. Lebedev Physical Institute was financially supported by the Ministry of Science and Higher Education of the Russian Federation, Contract No. 075-15-2021-598. E.V.K. and M.S.K. acknowledge the Saint-Petersburg State University (Grant No. 122040800257-5).

%

\textbf{AUTHOR INFORMATION}

{\bf Corresponding Authors} \\
Sergey~R.~Meliakov,  Email: melyakovs@lebedev.ru   \\
Dmitri R. Yakovlev,  Email: dmitri.yakovlev@tu-dortmund.de\\

\textbf{ORCID}\\
Sergey~R.~Meliakov         0000-0003-3277-9357 \\  
Evgeny~A.~Zhukov:          0000-0003-0695-0093 \\  
Vasilii~V.~Belykh:          0000-0002-0032-748X \\ 
Mikhail~O.~Nestoklon:      0000-0002-0454-342X  \\ 
Elena V. Kolobkova:        0000-0002-0134-8434 \\  
Maria S. Kuznetsova:       0000-0003-3836-1250 \\  
Manfred~Bayer:             0000-0002-0893-5949 \\ 
Dmitri R. Yakovlev:        0000-0001-7349-2745 \\  


\section*{References}


\begin{thebibliography}{}

\bibitem{Kovalenko2017}
M. V. Kovalenko, L. Protesescu, and M. I. Bondarchuk,
Properties and potential optoelectronic applications of lead halide perovskite nanocrystals,
\textit{Science} \textbf{2017}, {358}, 745--750.	


\bibitem{Shamsi2019} J. Shamsi, A. S. Urban, M. Imran, L. De Trizio, and L. Manna,
Metal halide perovskite nanocrystals: synthesis, post-synthesis modifications, and their optical properties,
\textit{Chemical Reviews} \textbf{2019}, {119}, 3296--3348.

\bibitem{Dey2021} A. Dey, J. Ye, A. De, E. Debroye, S. K. Ha, et al.,
State of the art and prospects for halide perovskite nanocrystals,
\textit{ACS Nano} \textbf{2021}, {15}, 10775--10981.

\bibitem{Mu2023} Y. Mu, Z. He, K. Wang, X. Pi, and S. Zhou,
Recent progress and future prospects on halide perovskite nanocrystals for optoelectronics and beyond,
\textit{iScience} \textbf{2023}, {25}, 105371.

\bibitem{Huang2023} C.-Y. Huang, H. Li, Y. Wu, C.-H. Lin, X. Guan, L. Hu, J. Kim, X. Zhu, H. Zeng, and T. Wu,
Inorganic halide perovskite quantum dots: A versatile nanomaterial platform for electronic applications,
\textit{Nano-Micro Lett.} \textbf{2023}, {15}, 16.

\bibitem{Li2017}
P. Li, C. Hu, L. Zhou, J. Jiang, Y. Cheng, M. He, X. Liang, and W. Xiang,
Novel synthesis and optical characterization of CsPb$_2$Br$_5$ quantum dots in borosilicate glasses,
\textit{Mater. Lett.} \textbf{2017}, {209}, 483--485.

\bibitem{Liu2018}
S. Liu, Y. Luo, M. He, X. Liang, and W. Xiang,
Novel CsPbI$_3$ QDs glass with chemical stability and optical properties,
\textit{J. Eur. Ceram. Soc.} \textbf{2018}, {38}, 1998--2004.

\bibitem{Liu2018a}
S. Liu, M. He, X. Di, P. Li, W. Xiang, and X. Liang,
Precipitation and tunable emission of cesium lead halide perovskites (CsPbX$_3$, X = Br, I) QDs in borosilicate glass,
\textit{Ceram. Int.} \textbf{2018}, {44}, 4496--4499.

\bibitem{Ye2019}
Y. Ye, W. Zhang, Z. Zhao, J. Wang, C. Liu, Z. Deng, X. Zhao, and J. Han,
Highly luminescent cesium lead halide perovskite nanocrystals stabilized in glasses for light-emitting applications,
\textit{Adv. Opt. Mater.} \textbf{2019}, {7}, 1801663.

\bibitem{Kolobkova2021}
E. V. Kolobkova, M. S. Kuznetsova,  and N. V. Nikonorov,
Perovskite CsPbX$_3$ (X = Cl, Br, I) nanocrystals in fluorophosphate glasses,
\textit{J. Non-Crystalline Solids} \textbf{2021}, {563}, 120811.

\bibitem{Belykh2022} V. V. Belykh,  M. L.  Skorikov, E. V. Kulebyakina,  E. V. Kolobkova, M. S.  Kuznetsova,  M. M. Glazov, and D. R. Yakovlev,
Submillisecond spin relaxation in CsPb(Cl,Br)$_3$ perovskite nanocrystals in a glass matrix,
\textit{Nano Lett.}  \textbf{2022}, {22}, 4583--4588.

\bibitem{Efros2021} Al. L. Efros, and L. E. Brus,
Nanocrystal quantum dots: From discovery to modern development,
	\textit{ACS Nano} \textbf{2021}, {15}, 6192--6210.

\bibitem{kirstein2022nc}
E.~Kirstein, D.~R.~Yakovlev, M.~M.~Glazov, E.~A.~Zhukov, D.~Kudlacik, I.~V.~Kalitukha, V.~F.~Sapega, G.~S.~Dimitriev, M.~A.~Semina, M.~O.~Nestoklon, E.~L.~Ivchenko, N.~E.~Kopteva, D.~N.~Dirin, O.~Nazarenko, M.~V.~Kovalenko, A.~Baumann, J.~H\"ocker, V.~Dyakonov, and M.~Bayer,
The Land\'e factors of electrons and holes in lead halide perovskites: universal dependence on the band gap,
\textit{Nat. Commun.} {\textbf{2022}}, {13}, {3062}.

\bibitem{nestoklon2023_nl}
M. O. Nestoklon, E. Kirstein, D. R. Yakovlev, E. A. Zhukov, M. M. Glazov, M. A. Semina, E. L. Ivchenko, E. V. Kolobkova, M. S. Kuznetsova, and M. Bayer,
Tailoring the electron and hole Land\'e factors in lead halide perovskite nanocrystals by quantum confinement and halide exchange,
\textit{Nano Lett.} \textbf{2023}, {23}, 8218--8224.

\bibitem{Vardeny2022_book} \textit{Hybrid Organic Inorganic Perovskites: Physical Properties and Applications}. (eds. Z.~V. Vardeny,  M.~C. Beard) (World Scientific, \textbf{2022}). Volume 3: Spin Response of Hybrid Organic Inorganic Perovskites.

\bibitem{Nestoklon2018}
M. O. Nestoklon,  S. V.  Goupalov, R. I.  Dzhioev,  O. S.  Ken, V. L. Korenev,  Yu. G.  Kusrayev, V. F.   Sapega,  C.  de Weerd,  L.  Gomez,  T.  Gregorkiewicz,  J.  Lin,  K.  Suenaga,  Y.  Fujiwara,  L. B.  Matyushkin, and I. N. Yassievich,
 Optical orientation and alignment of excitons in ensembles of inorganic perovskite nanocrystals,
\textit{Phys. Rev. B}  \textbf{2018}, {97}, 235304.

\bibitem{canneson2017}
D.~Canneson, E.~V. Shornikova, D.~R. Yakovlev, T.~Rogge, A.~A. Mitioglu, M.~V. Ballottin, P.~C.~M. Christianen, E.~Lhuillier, M.~Bayer, and L.~Biadala,
Negatively charged and dark excitons in CsPbBr$_3$ perovskite nanocrystals revealed by high magnetic fields,
\textit{Nano Lett.}  \textbf{2017}, {17}, 6177--6183.


\bibitem{Crane2020} M. J. Crane,  L. M.  Jacoby,  T. A.  Cohen,  Y. Huang,  C. K. Luscombe, and D. R. Gamelin,
Coherent spin precession and lifetime-limited spin dephasing in CsPbBr$_3$ perovskite nanocrystals,
\textit{Nano Lett.}  {\textbf{2020}}, {20}, 8626--8633.

\bibitem{Grigoryev2021} P. S. Grigoryev,  V. V.  Belykh,  D. R.  Yakovlev,  E. Lhuillier, and M. Bayer,
Coherent spin dynamics of electrons and holes in CsPbBr$_3$ colloidal nanocrystals,
\textit{Nano Lett.}  {\textbf{2021}}, {21}, 8481--8487.

\bibitem{Lin2022} X. Lin,  Y.  Han, J.  Zhu, and K. Wu,
Room-temperature coherent optical manipulation of hole spins in solution-grown perovskite quantum dots,
\textit{Nat. Nanotechnol.}  {\textbf{2022}}, {18}, 124.

\bibitem{Meliakov2023NCs} S. R. Meliakov, E. A. Zhukov, E. V. Kulebyakina,  V. V. Belykh, and D. R. Yakovlev,
Coherent spin dynamics of electrons in CsPbBr$_3$ perovskite nanocrystals at room temperature,
\textit{Nanomaterials} \textbf{2023}, {13}, 2454.

\bibitem{kirstein2023_SML}
E. Kirstein, N.~E. Kopteva, D.~R. Yakovlev, E.~A. Zhukov, E.~V. Kolobkova, M.~S. Kuznetsova, V.~V. Belykh, I.~A. Yugova, M.~M. Glazov, M.~Bayer, and A. Greilich,
Mode locking of hole spin coherences in CsPb(Cl,Br)$_3$ perovskite nanocrystals,
\textit{Nat. Commun.}  {\textbf{2023}}, {14}, 699.

\bibitem{Strohmair2020} S. Strohmair,  A. Dey, Y. Tong, L. Polavarapu, B. J.  Bohn, and J. Feldmann,
Spin polarization dynamics of free charge carriers in CsPbI$_3$ nanocrystals,
\textit{Nano Lett.}  {\textbf{2020}},  {20}, 4724--4730.

\bibitem{Han2022}
Y. Han, W. Liang, X. Lin, Y. Li, F. Sun, F. Zhang, P. C. Sercel, and K. Wu, 
Lattice distortion inducing exciton splitting and coherent quantum beating in CsPbI$_3$ perovskite quantum dots,
\textit{Nature Materials} \textbf{2022}, {21}, 1282-1289.

\bibitem{Gao2024}
K. Gao, Y. Li, Y. Yang, Y. Liu, M. Liu, W. Liang, B. Zhang, L. Wang, J. Zhu, and K. Wu, 
Manipulating coherent exciton dynamics in CsPbI$_3$ perovskite quantum dots using magnetic field,
\textit{Adv. Mater.} \textbf{2024}, {36}, 2309420.

\bibitem{Zhu2024}
J. Zhu, Y. Li, X. Lin, Y. Han, and K. Wu, 
Coherent phenomena and dynamics of lead halide perovskite nanocrystals for quantum information technologies,
\textit{Nature Materials} \textbf{2024}, {23}, 1027-1040.


\bibitem{Kopteva_gX_2023} N. E. Kopteva,  D. R.  Yakovlev, E.  Kirstein,  E. A. Zhukov, D.  Kudlacik,   I. V. Kalitukha,  V. F. Sapega, D. N.  Dirin,  M. V. Kovalenko, A.  Baumann,  J. H\"ocker, V.  Dyakonov,  S. A. Crooker, and M. Bayer,
Weak dispersion of exciton Land\'e  factor with band gap energy in lead halide perovskites: Approximate compensation of the electron and hole dependences,
\textit{Small} \textbf{2023}, 2300935.

\bibitem{Meliakov2024paper1} S. R. Meliakov, E. A. Zhukov, V. V. Belykh, M. O. Nestoklon, E. V. Kolobkova, M. S. Kuznetsova, M. Bayer, and D. R. Yakovlev,
Temperature dependence of the electron and hole Land\'e $g$-factors in CsPbI$_3$ nanocrystals in a glass matrix,
https://arxiv.org/abs/2407.21610    placed 31.07.2024.

\bibitem{Meliakov2024paper3} S. R. Meliakov, V. V. Belykh, E. A. Zhukov, E. V. Kolobkova, M. S. Kuznetsova, M. Bayer, and D. R. Yakovlev,
Hole spin precession and dephasing induced by nuclear hyperfine fields in CsPbBr$_3$ and CsPb(Cl,Br)$_3$ nanocrystals in a glass matrix,
https://arxiv.org/abs/2409.01065     placed 02.09.2024.

\bibitem{Yakovlev_Ch6}
D.~R. Yakovlev and M. Bayer,
Coherent spin dynamics of carriers. In \textit {Spin Physics in Semiconductors}, M.~I. Dyakonov (ed.) (Springer International Publishing AG, \textbf{2017}) chapter 6, pp. 155--206.

\bibitem{belykh2019}
V. V. Belykh, D. R. Yakovlev, M. M. Glazov, P. S. Grigoryev, M. Hussain, J. Rautert, D. N. Dirin, M. V. Kovalenko, and M. Bayer,
Coherent spin dynamics of electrons and holes in CsPbBr$_3$ perovskite crystals,
\textit{Nat. Commun.} {\textbf{2019}}, {10}, 673.

\bibitem{YugovaPRB09} I. A. Yugova,  M. M. Glazov,  E. L. Ivchenko, and  Al. L. Efros,
Pump-probe Faraday rotation and ellipticity in an ensemble of singly charged quantum dots,
\textit{Phys. Rev. B} {\textbf{2009}}, {80}, {104436}.

\bibitem{Glazov2010} M. M. Glazov, I. A. Yugova, S. Spatzek, A. Schwan, S. Varwig, D. R. Yakovlev, D. Reuter, A. D. Wieck, and M. Bayer,
Effect of pump-probe detuning on the Faraday rotation and ellipticity signals of mode-locked spins in (In,Ga)As/GaAs quantum dots,
\textit{Phys. Rev. B} \textbf{2010}, {82}, 155325.

\bibitem{Nestoklon2021}
M. O. Nestoklon, 
Tight-binding description of inorganic lead halide perovskites in cubic phase, 
\textit{Comput. Mater. Science} \textbf{2021}, {196}, 110535.

\bibitem{Kashikar2018}
R. Kashikar, B. Khamari, and B. R. K. Nanda, 
Second-neighbor electron hopping and pressure induced topological quantum phase transition in insulating cubic perovskites,
\textit{Phys. Rev. Mater.} \textbf{2018}, {2}, 124204.


\end{thebibliography}
\end{document}